**Spread of Covid-19 in urban neighbourhoods and slums of the developing world**

Anand Sahasranaman[1,2,*] and Henrik Jeldtoft Jensen[2,3,#]

[1]Division of Sciences and Division of Social Sciences, Krea University, Sri City, AP 517646, India

[2]Centre for Complexity Science and Dept. of Mathematics, Imperial College London, London SW72AZ, UK.

[3]Institute of Innovative Research, Tokyo Institute of Technology, 4259, Nagatsuta-cho, Yokohama 226-8502, Japan.

[*] Corresponding Author. Email: anand.sahasranaman@krea.edu.in

[#] Email: h.jensen@imperial.ac.uk

**Abstract:**

We study the spread of Covid-19 across neighbourhoods of cities in the developing world and find that small numbers of neighbourhoods account for a majority of cases ($k$-index$\sim$0.7). We also find that the countrywide distribution of cases across states/provinces in these nations also displays similar inequality, indicating self-similarity across scales. Neighbourhoods with slums are found to contain the highest density of cases across all cities under consideration, revealing that slums constitute the most at-risk urban locations in this epidemic. We present a stochastic network model to study the spread of a respiratory epidemic through physically proximate and accidental daily human contacts in a city, and simulate outcomes for a city with two kinds of neighbourhoods - slum and non-slum. The model reproduces observed empirical outcomes for a broad set of parameter values - reflecting the potential validity of these findings for epidemic spread in general, especially across cities of the developing world. We also find that distribution of cases becomes less unequal as the epidemic runs its course, and that both peak and cumulative caseloads are worse for slum neighbourhoods than non-slums at the end of an epidemic. Large slums in the developing world therefore contain the most vulnerable populations in an outbreak, and the continuing growth of metropolises in Asia and Africa presents significant challenges for future respiratory outbreaks from perspectives of public health and socioeconomic equity.

## 1. Introduction:

In the wake of the novel coronavirus Covid-19 pandemic that is currently sweeping the planet, there is increasing concern over the impact on large urban slums in the developing world. This concern primarily stems from the nature of dwelling arrangements in developing cities, where large proportions of the population live in densely populated slums and shantytowns [1]. Broadly, slums are defined as "communities characterized by insecure residential status, poor structural quality of housing, overcrowding, and inadequate access to safe water, sanitation, and other infrastructure" [2]. This definition emphasizes the fact that slums house the poorest and most vulnerable populations in cities. UN-Habitat estimated that over a third of the urban population lived in slums in 2012-13, with significant geographical heterogeneity – the proportion was 62% for sub-Saharan Africa and 35% for southern Asia, and 25% for Latin America [1]. The sheer scale of slums is further exacerbated by the density of population in such settlements. Table 1 presents some statistics on the density of living in some of the largest metropolises of the developing world and shows that these cities have high average population densities (and high slum populations), but individual slum neighbourhoods even within these cities often show population densities an order of magnitude higher, suggesting significant intra-city heterogeneity in densities of living.

| City | Slum population (%) | Average population density (per $km^2$) | Large slum | Slum population density (per $km^2$) |
|---|---|---|---|---|
| Mumbai, India | 41% | 25,771 | Dharavi | 335,900 |
| Cape Town, South Africa | 35% | 1,520 | Khayelishta | 10,120 |
| Rio de Janeiro, Brazil | 22% | 5,231 | Mare | 30,400 |
| Dhaka, Bangladesh | 38% | 19,501 | Korail | 205,410 |
| Lagos, Nigeria | 70% | 18,788 | Mushin | 128,882 |
| Manila, Philippines | 31% | 44,866 | Tondo | 73,548 |

**Table 1**: *Densities of living in developing world metropolises and their slum neighbourhoods*.

It is important to remember that the high population densities in developing cities are being attained without building vertically (unlike cities like New York City, Seoul, or Tokyo), with typical living conditions in slums described as small single room shacks ($\sim 10\ m^2$) with around 5 people living in them, situated adjacent to one another, and with up to 10 families sharing a water tap and a pit latrine [3]. High population density achieved under such conditions therefore creates an environment rife for epidemic spread through air or water. Our specific concern relates to the spread of disease through such urban slums, which represent a critical feature of urbanization in developing nations [1], especially in the context of infectious disease outbreaks like Covid-19 where viral transmission is aided by increased population density, manifested as more frequent person-to-person contact, crowded housing, and unsanitary environments [4, 5].

In this work, we use Covid-19 caseload data at a sub-city level (ward or neighbourhood or local government level) to empirically characterize the spread of the epidemic across urban neighbourhoods in six developing world metropolises, specifically to understand the nature of infectious spread at fine-grained levels in contexts where slums are a salient feature of the urban landscape. Based on this characterization, we explore the differential impacts of Covid-19 across slum and non-slum neighbourhoods in these cities. In order to explore whether the observed epidemy outcomes could be representative of other similar epidemics in developing urban contexts, we create a network model to propagate the stochastic dynamics of the spread of infection through an urban system (city) and estimate the differential impacts on slum and non-slum populations. Finally, we discuss the results obtained in the context of cities in the developing world.

## 2. Evidence on impact of Covid-19 on cities and slums:

We focus our attention on six specific cities (Table 1) because they are amongst the primate cities of the global south; are severely impacted by Covid-19; and have made available data at the required level of local granularity to enable this fine-grained analysis. However, even for many of these cities, data at the sub-city level is not released regularly and is only available occasionally. We discuss all sources of data and constraints in Appendix 1. Table 2 provides greater detail on the sub-city units we consider for the analysis.

| City | Nature of sub-city unit | Number of sub-city units | Average population of sub-city unit | Caseload as of (date) | Total caseload |
|---|---|---|---|---|---|
| Mumbai Corporation | Ward | 24 | 518,432 | July 20, 2020 | 99,566 |
| City of Cape Town | Suburb, Township | 58 | 64,483 | June 22, 2020 | 38,540 |
| Rio de Janeiro | Bairro | 163 | 40,261 | July 21, 2020 | 61,818 |
| Dhaka City | Thana | 41 | 161,711 | June 28, 2020 | 15,754 |
| Lagos Metropolitan Area | Local Govt. Area (LGA) | 16 | 1,375,405 | May 7, 2020 | 1,352 |
| City of Manila | District | 16 | 110,429 | July 10, 2020 | 3,248 |

**Table 2**: *Data details.*

We first study the distribution of cumulative caseloads across sub-city units (we will refer to these sub-city units generally as neighbourhoods) for each of the six cities and find that cases show an unequal distribution across neighbourhoods, with a high proportion of cases contained in a small proportion of neighbourhoods - the top 20% of neighbourhoods, in terms of Covid-19 caseloads, account for 31% of cases in Mumbai, 69% in Cape Town, 58% in Rio de Janeiro, 50% in Dhaka, 65% in Lagos, and 55% in Manila respectively (Fig. 1, black). The emergence of such a relationship across neighbourhoods in all cities under consideration - given the underlying heterogeneity in terms of numbers of sub-city units, population scale of units, and total caseload - suggests that the outcome is robust and representative of real underlying dynamics of infectious spread. We characterize the unequal nature of this spread across neighbourhoods using the $k$-index, which is a measure of inequality that generalizes Pareto's 80-20 rule – that is, given the cumulative distribution of an attribute (such as Covid-19 caseload) across a set of entities (such as neighbourhoods), the $k$-index has the property that $(1 - k_f)$ proportion of entities contain $k_f$ proportion of the attribute [6]. We find that, apart from Mumbai ($k_f = 0.57$), all other cities have much higher $k_f \sim 0.70$ ($k_f$ for Cape Town, Rio de Janeiro, Dhaka, Lagos, and Manila are $0.75, 0.70, 0.68, 0.74, 0.70$ respectively), meaning that $\sim 30\%$ of the neighbourhoods in these cities account for $\sim 70\%$ of reportedcases. We also study the time evolution of the distribution of cases in these cities, considering two points in time that are around a month apart (subject to data availability as highlighted in Appendix 1), and find that the $k$-index of the distribution appears to decrease over time for most cities (Mumbai, Cape Town, and Lagos), while it remains consistent for Rio de Janeiro and Dhaka, and marginally increases for Manila (Fig. 1, red).

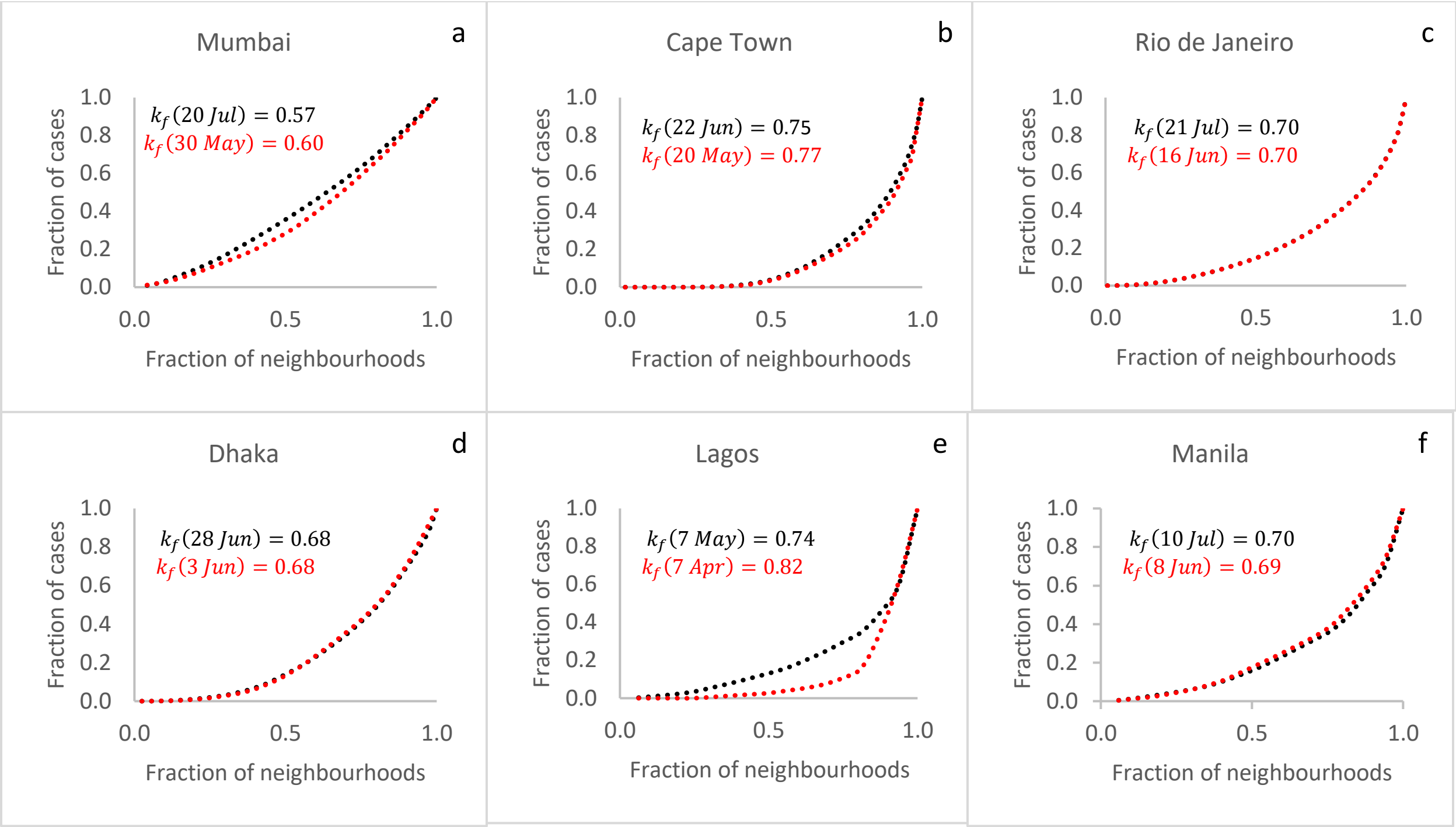

**Figure 1**. *Distribution of Covid-19 cases across neighbourhoods*. Fraction of cases v. Fraction of neighbourhoods. a: Mumbai. b: Cape Town. c: Rio de Janeiro. d: Dhaka. e: Lagos. f: Manila. Black dots: Most recent caseload distribution, dates as per Table 2. Red dots: Older caseload distributions – Mumbai (May 30), Cape Town (May 20), Rio de Janeiro (June 16), Dhaka (June 3), Lagos (April 7), and Manila (June 8). The $k$-index appears to decline over time.

When we explore the distribution of Covid-19 caseload across states or provinces within the countries containing these six cities, we again find that $k_f \sim 0.70$, which is very similar to the $k$-index observed for caseload distribution within these cities. The distribution of cases across the states of India, states of Brazil, states of Nigeria, provinces of South Africa, and districts of Bangladesh yield $k$-indices of $0.77, 0.65, 0.75, 0.70,$ and $0.76$ respectively (we were unable to find province level data for Philippines). Therefore, the distribution of caseload across states/provinces in nations mirrors the distribution across neighbourhoods in cities, indicating self-similar behaviour across scales.

Given this unequal distribution, we now explore the characteristics of neighbourhoods that have the highest caseloads. Our current understanding of Covid-19 suggests that physical proximity is an important determinant of local spread. Therefore, we study caseloads across neighbourhoods in all six cities, with a focus on differential impacts of Covid-19 on high density neighbourhoods with slums, and other neighbourhoods. In order to do this, we first map large slum settlements in these cities to the appropriate sub-city unit; it is important to note that given the high levels of slum population in these cities (and, generally, in cities across the developing world) almost all neighbourhoods have pockets of slums, so we label as ‘neighbourhoods with slums’ those sub-city units which show a high concentration of slums as revealed by slum mapping exercises (detailed in Appendix 2). The resulting list of neighbourhoods with slums across the 6 cities is: 11 out of 23 wards in Mumbai (G-North - containing the Dharavi slum, G-South, F-South, L, N, H-East, M-East, M-West, K-East, K-West, P-North); 8 out of 58 suburbs/townships in Cape Town (Khayelitsha, Mitchells Plain, Gugulethu, Delft, Philippi, Nyanga, Langa, Mfuleni); 41 out of 163 bairros in Rio de Janeiro (Rocinha, Jacarezihno, Mare, Cidade de Deus, Complexo do Alemao, Mangueira, Penha, Acari, Tijuca, Costa Barros, Ramos, Benfica, Pavuna, Encantado, Lins de Vasconcelos, Manguinhos, Madureira, Inhaumos, Rio Comprido, Iraja, Anchieta, Vigario Geral, Guadalupe, Cordovil, Piedade, Jacare, Parada de Lucas, Copacabana, Tomas Coelho, Magalhaes Bastos, Realengo, Bangu, Jacarepagua, Andarai, Bras de Pina, Honorio Gurgel, Engenho Novo, Turiacu, Padre Miguel, Coelho Neto, Engenho de Dentro); 12 out of 41 thanas in Dhaka (Mirpur, Gulshan - containing the Korail slum, Mohammadpur, Jatrabari, Lalbagh, Sutrapur, Chak Bazar, Gendaria, Hazaribagh, Kotwali, Kamrangir Char, Shyampur); 5 out of 16 LGAs in Lagos (Agege, Ajeromi-Ifelodun, Mushin, Somolu, Lagos Island, and Lagos Mainland - containing the floating Makoko slum); and 2 out of 17 districts in Manila (Tondo and San Andres).

We find, in line with expectations, that average population densities of neighbourhoods with slums are much higher than other neighbourhoods (Fig. 2a, neighbourhoods with slums – red, other neighbourhoods - blue). When we assess the distribution of cases across neighbourhood types, taking into account density of living, we find that caseload per capita, represented by caseload per million population (Fig. 2b), and caseload per unit area ($km^2$) (Fig. 2c) are systematically higher in neighbourhoods with slums than in non-slum neighbourhoods across cities.

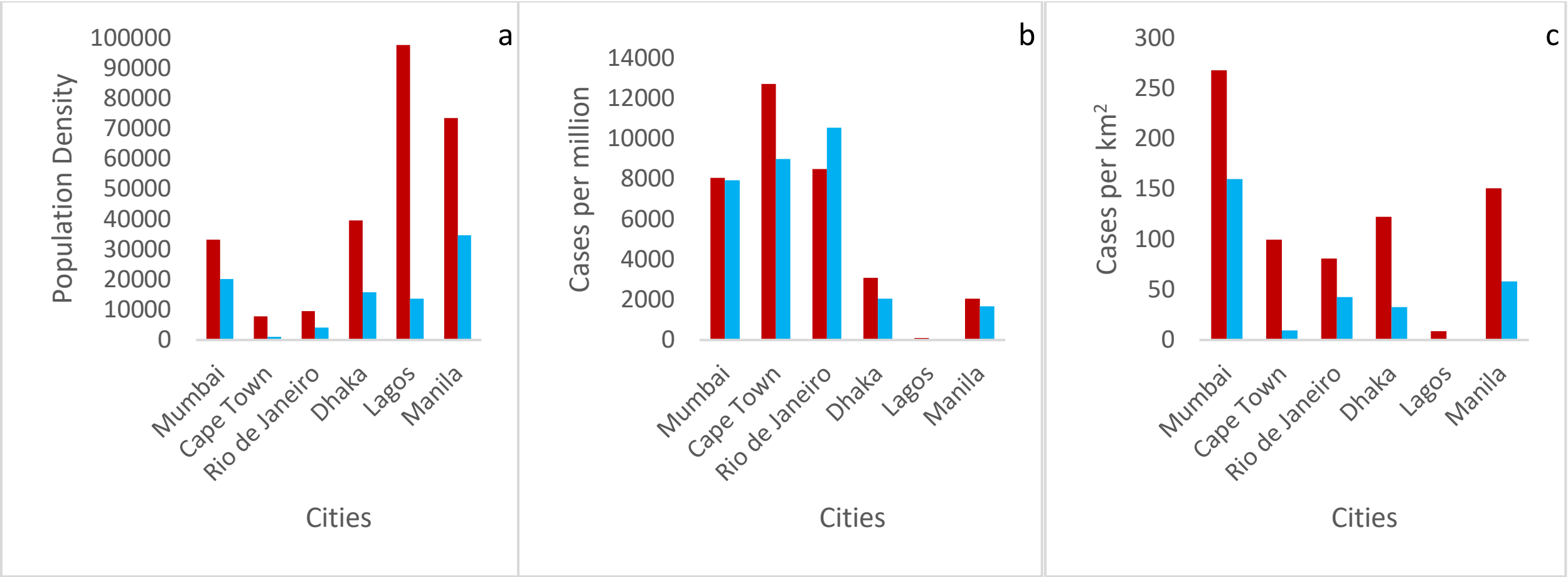

**Figure 2**. *Distribution of Covid-19 cases across neighbourhoods with slums and other neighbourhoods*. Red columns: Neighbourhoods with slums. Blue columns: Non-slum neighbourhoods. a: Population density of slum and non-slum neighbourhoods shows that slum neighbourhoods have higher densities, on average, across all cities. b: Covid-19 cases per million population across slum and non-slum neighbourhoods. Slum neighbourhoods are more affected in all cities, except Rio de Janeiro. c: Covid-19 cases per $km^2$ across slum and non-slum neighbourhoods. In all cities, slum neighbourhoods show much higher spatial density of cases than non-slum neighbourhoods.

The only exception here is Rio de Janeiro, where we find that neighbourhoods with slums have a lower caseload per capita than non-slum neighbourhoods; this should however be seen in light of the many concerns have been raised about testing and measurement of Covid-19 cases in Brazil's favelas [7, 8, 9].

Overall, this finding suggests that density of living in neighbourhoods is a critical mediator of the dynamics of infectious spread in a city, and that the urban poor in slums are starkly worse off in terms of epidemy outcomes. Given this empirical context, we attempt to model these differential impacts by simulating the spread of a typical respiratory epidemic in a network representing a city-system composed of slum and non-slum neighbourhoods, and test whether the observed intra-city epidemic outcomes from Covid-19 are representative more generally of respiratory epidemic spread in cities with slums.

## 3. Model definition and specifications:

We model a network of a city consisting of $N$ nodes, with each node representing an agent in the city, and $H$ neighbourhoods amongst which agents are distributed. While we lack empirical data on the structure of real networks of physical proximity in cities of the developing world, there is a growing body of work indicating that highly connected nodes or 'super-spreaders' are disproportionately important in the transmission of even influenza like illnesses [10, 11, 12]. Therefore, we propose to explore the dynamics of transmission on a scale-free Barabási-Albert (BA) network [13]. The Barabási-Albert (BA) network with $N$ nodes is generated by attaching new nodes with $m$ neighbours, such that the links of a new node show preferential attachment for existing nodes with high degree. We also test the robustness of model outcomes for sensitivity to network type in Appendix 3.

Our interest is in studying differential impacts across slum and non-slum neighbourhoods described by a wide variation in population density, and we simulate such density differences as differences in the average degree of nodes in each neighbourhood of the network. That is, we model connectedness of a neighbourhood as the average degree of nodes in a neighbourhood. We construct neighbourhoods in the network by ordering all $N$ agents based on node degree and then allotting each of them in order to the $H$ neighbourhoods in the city system, such that the first neighbourhood is filled with the first set of ordered $N/H$ agents, followed by the second neighbourhood and so on, until the final neighbourhood is filled with the last set of $N/H$ agents. Creating neighbourhoods in this way ensures that average degree of nodes across neighbourhoods shows significant heterogeneity. It also means that nodes in neighbourhoods with higher average degree are connected to many nodes both within and outside of their neighbourhoods – this is meaningfully representative of the urban poor in cities of the developing world, who live in densely populated slums and have high interconnectedness (unavoidable physical proximity) within the slum, but work largely in other non-slum neighbourhoods, including as essential services workers such as sanitation and health workers. This algorithm also means that neighbourhoods with high average degree correspond only to high-density slum neighbourhoods, and not high-density neighbourhoods in general – for instance, neighbourhoods that well-off and where high densities are obtained by building vertically are represented in our model as neighbourhoods with lower average node degree (lower connectedness), which is a more likely representation of their daily contact networks. We also vary the algorithm to populate neighbourhoods, while still ensuring reasonable heterogeneity in average degree of neighbourhoods, and find that model outcomes are robust to these changes (Appendix 3).

We use the three compartment Susceptible-Infected-Recovered (SIR) model as the basis for an agent's progression through the duration of the epidemic [14]. Agents start out in the Susceptible ($S$) compartment until the time they are infected, at which point they fall into the Infected ($I$) compartment. After spending a specified duration of time being infected, when they spread the infection in the network, they move to the Recovered ($R$) compartment, at which time they are immune - neither infective nor susceptible to the infection again. At $t = 0$ days, we have one random node that is infected ($I$), while the remaining $N - 1$ nodes are susceptible ($S$).

At each time step $t$, the dynamics of infectious spread in the network are modelled as follows: First, each infected ($I$) agent spreads the disease to each of its susceptible ($S$) neighbours in the network with transmission probability $p$. Given a node $i$ with $k$ neighbours (or a contact rate of $k$), the average daily infections caused by this node, or its daily transmission rate ($\beta_i$), is the product of the transmission probability and the contact rate of the node, $\beta_i = pk$. Second, each infected agent moves into the recovered ($R$) compartment if it has spent $1/\gamma$ days in the infected ($I$) compartment. $\gamma$ is defined to be the recovery rate and remains constant through the dynamics.

We propagate these dynamics over a period of $t = T_f$ days and study the distribution of cases across the $H$ neighbourhoods over time, as well as the current and cumulative caseloads across neighbourhoods over time. Table 3 provides the complete set of parameter values and initial conditions for the simulations.

| | **Values** |
|---|---|
| **Parameters** | |
| Population – number of network nodes, $N$ | 10,000 |
| Number of edges from new node to extant nodes, $m$ (for Barabási-Albert graph) | 50 |
| Number of neighbourhoods, $H$ | 20 |
| Probability of node to node transmission, $p$ | 0.004 |
| Number of iterations (days) in one simulation of model, $T_f$ | 120 |
| Number of simulations | 100 |
| **Initial Conditions** | |
| Number of susceptible nodes, $S(0)$ | 9,999 |
| Number of infectious nodes, $I(0)$ | 1 |
| Number of recovered nodes, $R(0)$ | 0 |

**Table 3**: *Parameter values and initial conditions.*

We also propagate model dynamics for a range of parameter values: population ($N$), neighbourhoods ($H$), probability of transmission ($p$), as well as for varying network type and mechanisms for populating neighbourhoods. The results are presented in Appendix 3.

## 4. Results:

The evolution of cumulative fraction of caseload across neighbourhoods clearly shows that the rate of case growth increases with population density (Fig. 3a). This is in keeping with the empirical finding that once epidemy dynamics are underway and the infection has reached higher density neighbourhoods, caseload per capita is higher in high density neighbourhoods. For instance, at day 10 of the dynamics, the densest neighbourhood in our network (with average node degree, $k = 421$) has a cumulative caseload of 4.7% (as a fraction of its population), while the lowest density neighbourhood (with average node degree, $k = 50$) is at 0.5%, and all other neighbourhoods with densities in between these extremes show caseloads between 0.5% and 4.7% (Fig. 3a). The corresponding caseloads on days 20 and 30 are 66% and 96% for the densest neighbourhood, and 24% and 71% for the lowest density neighbourhoods.

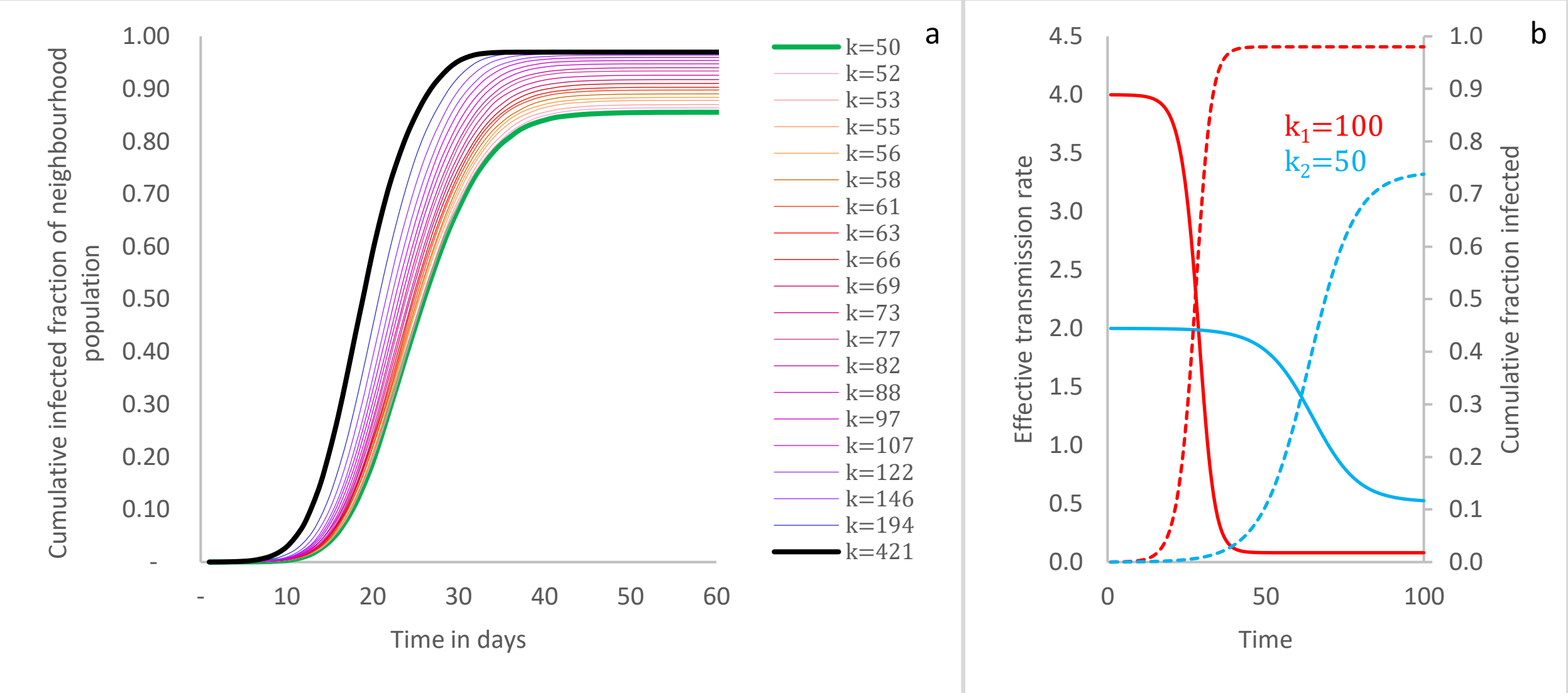

**Figure 3**. *Cumulative caseload simulation and analytical description*. a: Simulation of cumulative infected population fractions across neighbourhoods over time. Higher density neighbourhoods not only show earlier and steeper rise in caseloads but also end up with higher fractions of populations infected at the end of the epidemic. b: Effective transmission rate (and cumulative fraction infected) over time. Analytical description of the evolution of effective transmission rates for neighbourhoods with different average contact rates ($k$). Denser neighbourhood is represented by $k_1 = 100$ and spearser neighbourhood by $k_2 = 50$.

To explore these dynamics analytically, consider a neighbourhood with $N_h$ nodes, each with degree $k$. Given $p$ and $k$, at $t = 0$, the average daily transmission rate is $\beta = pk$. At the end of a time interval $t$, let $f_S(t)$ be the fraction of population still susceptible and $f_I(t)$ the fraction that has ever been infected until $t$, such that $f_S(t) = 1 - f_I(t)$. $f_I(t)$ is given by (Eq. 1):

$$f_I(t) = \begin{cases} \frac{1}{N_h}, & at\ t = 0 \\ f_I(t-1) + (\beta f_I(t-1)\big(1 - f_I(t-1)\big), & for\ 1 \le t < 1/\gamma \\ f_I(t-1) + (\beta \left( f_I(t-1) - f_I\left(t - \frac{1}{\gamma}\right)\right)\big(1 - f_I(t-1)\big), & for\ t \ge 1/\gamma \end{cases} \quad (1)$$

The effective transmission rate of the epidemic in the neighbourhood, $R_e(t)$, is the average number of people infected by an individual in the neighbourhood at time $t$:

$$R_e(t) = \frac{pf_S(t-1)k}{\gamma} = \beta f_S(t-1)/\gamma \quad (2)$$

Using this simple construct, we consider two neighbourhoods – a slum with average degree $k_1$ and a non-slum with average degree $k_2$ ($k_2 < k_1$) - with probability of transmission $p$ and a single node infected at $t = 0$. The evolution of $R_e(t)$ shows that the slum has a much higher effective transmission rate in the early part of the dynamics due to higher $k$ (Fig. 3b). This results in sharp increase in caseloads in this period, causing a simultaneous sharp decline in $R_e(t)$ due to the coevolution of susceptible and infected populations. The non-slum neighbourhood has a lower effective transmission rate to begin with and shows a more gradual increase in cases. The overall effect is that higher density results in higher caseloads per capita in the slum as against the non-slum (Fig. 3b), which offers a possible explanation for the empirical observations from developing world cities where case density increases with neighbourhood population density (Figs. 2b and 2c).

We also study the distribution of cases across neighbourhoods and find that, just as observed empirically, there is an unequal distribution of caseload across neighbourhoods during the dynamics (Fig. 4a). However, as the epidemic runs its course, the inequality in distribution progressively reduces – Fig. 4a plots the distribution of caseloads at different points in time and we see that inequality in distribution of cases is greatest at $t = 10$ when $k_f = 0.62$, following which there is continuous reduction in inequality until $t = 50$ when $k_f = 0.51$, at which point the epidemic has ended. For the epidemic to end, it infects as much of the population as is required for the effective transmission rate to summarily decline below 1; therefore, even as dense slum neighbourhoods see their caseloads rise steeper and peak earlier (Fig. 4b), thus yielding higher inequality in case distribution, lower density non-slums are not immune to the epidemy and will see delayed but increasing caseloads resulting in declining inequality in the distribution towards the end of the epidemic (Figs. 3a, 3b, and 4a). Our empirical findings from Mumbai, Cape Town, and Lagos conform with this modelled outcome, though Rio de Janeiro, Dhaka and Manila do not (Fig. 1).

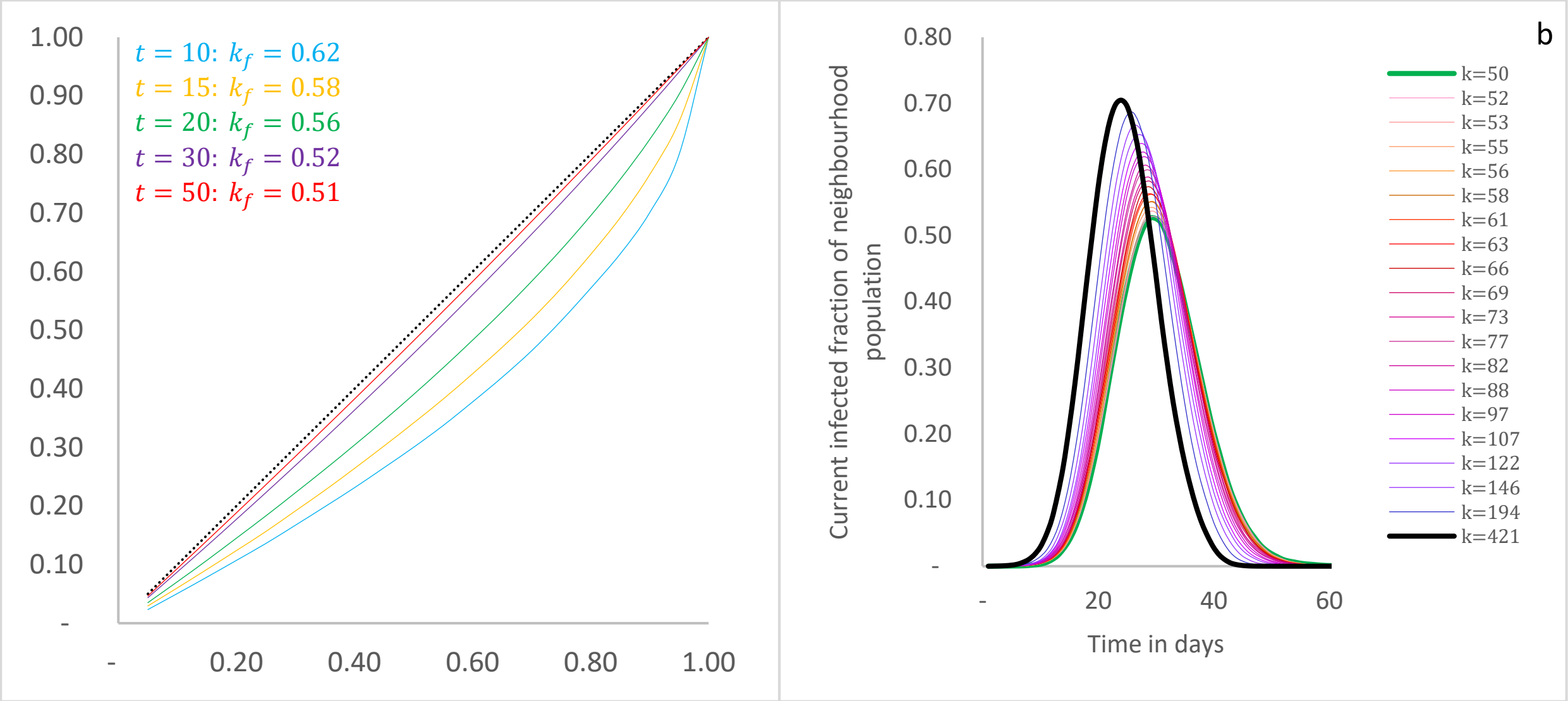

**Figure 4**. *Distribution and peak caseload simulation*. a: Simulation of distribution of cumulative caseloads across neighbourhoods over time. Distribution of cases is more unequal earlier in the epidemic and become less unequal as epidemic progresses b: Simulation of current caseloads across neighbourhoods over time. Higher density neighbourhoods show earlier and sharper peaks in active cases compared to lower density neighbourhoods. Dashed black line: Line of equality ($y = x$).

Our model suggests that both in terms of cumulative caseload outcomes at the end of the epidemic (Fig. 3a), as well as (higher and earlier) peak caseloads during the epidemic (Fig. 4b), slum neighbourhoods are much worse off than non-slums in an epidemic. We find that the nature of outcomes described here are robust to a wide range of change changes in model parameters such as population of the city system ($N$), probability of transmission ($p$), and number of neighbourhoods ($H$), as well as changes in network structure and mechanism of populating neighbourhoods (Appendix 3).

In summary, our modelled outcomes are in broad agreement with the empirical observations, suggesting that the nature of these outcomes is more generally reflective of epidemic spread in cities with slums.

## 5. Conclusion:

We study the evolution of the Covid-19 epidemic across neighbourhoods within a city, for a set of metropolises in the developing world. We find an unequal distribution of cases, with a small number of the most densely populated neighbourhoods containing a significant proportion of total caseload across all cities, as illustrated by a $k$-index $\sim 0.70$ across these cities. This finding appears to hold across scales, with national case distribution across these states/provinces also displaying similar inequality in case distribution. We also find that neighbourhoods with the highest case densities – both in terms of population and area - contain the largest slums in these cities, and that consequently the urban poor in slums are at the highest risk in this epidemic.

Using a simple network model, we simulate the emergence of differential outcomes for slums and non-slums in a city. Model outcomes replicate both unequal distribution of cases as well as higher case densities in high-density neighbourhoods, suggesting that these outcomes are reflective of outbreaks in general for cities with slums. In addition, simulations also predict that as the epidemic progresses, distribution of cases across neighbourhoods becomes less unequal, and that both peak caseloads and cumulative caseloads are worse for slum neighbourhoods vis-à-vis non-slum neighbourhoods in cities.

Urban slums reflect increased demographic growth, migration, population densities, and poverty, which are the main processes found to be linked with prevalence of infectious diseases [15]. There is evidence to suggest that slum populations scale super-linearly with city size [16], meaning that larger cities have more than proportionally larger slums. It is anticipated that there will be over 40 megacities in the world by 2030 and most will be located in the developing world [17]. The evolution of larger slums and higher population densities will mean that slums will continue to at the forefront of epidemics, both in terms of public health and socioeconomic outcomes.

In the immediate or near term, health departments in developing countries must prepare specific guidelines for physical distancing in high density settlements that are clearly communicated and can be implemented by slum dwellers, so that their exposure risks are minimized [18]. However, the long-term solution to this lies in ensuring that slum settlements, which often house a large proportion of the urban population, are provided with functioning environmental infrastructure for piped running water and private sanitation, waste management, and electricity, in addition to basic health infrastructure such as primary healthcare facilities [19, 20, 21]. Creating cleaner, more sanitary environments will make it possible to more systematically counter the easy spread of infections.

**Author Contributions:** AS and HJJ conceived the methodology, wrote the analytical description, reviewed and edited the manuscript. AS conceived the research, collected the data, performed the analysis, programmed the simulation and wrote the draft manuscript**.**

**Competing Interests:** The authors declare that they have no competing interests.

**Funding:** The authors received no funding for this work.